# Negative refraction based on purely imaginary conjugate metamaterials

**YANGYANG FU,**[1, 2] **YADONG XU,**[2,3] **AND HUANYANG CHEN**[1, 4]

[1] *Institute of Electromagnetics and Acoustics and Department of Electronic Science, Xiamen University, Xiamen 361005, China.*
[2] *College of Physics, Optoelectronics and Energy, Soochow University, No.1 Shizi Street, Suzhou 215006, China.*
[3]*ydxu@suda.edu.cn*
[4]*kenyon@xmu.edu.cn*

**Abstract:** By introducing a new mechanism based on purely imaginary conjugate metamaterials (PICMs), we reveal that bidirectional negative refraction and planar focusing can be obtained using a pair of PICMs, which is a breakthrough to the unidirectional limit in parity time (PT) symmetric systems. Compared with PT symmetric systems that require two different kinds of materials, the proposed negative refraction can be realized with only two identical media. In addition, asymmetric excitation with bidirectional total transmission is observed in our PICM system. Therefore, a new way to realize negative refraction is presented, with more properties than those in PT symmetric systems.

## 1. Introduction

In optics, refraction of light at an interface of two different homogeneous media is a significant phenomenon, playing important roles in many optical devices. Conventionally, for two materials with positive indices, the refracted light, determined by Snell's law, shares the same direction with the incident one. It's thereby called as positive refraction. When one of the two materials is replaced by a negative one, the refracted light possesses an opposite direction with the incident one, i.e., negative refraction [1]. In the past few years, negative refraction has attracted much attention, because of its numerous intriguing applications including the perfect lens [2], the reversed Cerenkov radiation [3], and the negative Goos-Hänchen shift [4]. Various schemes to realize negative refraction have been proposed, such as metamaterials [5-8], photonic crystals [9, 10], gradient gratings [11, 12], phase conjugating surfaces [13, 14] and nonlinear optical films [15, 16]. All these endeavors have their own merits for negative refraction, but they turn out experimentally and theoretically to be imperfect, because their performances are drastically restricted by the inherent loss. Therefore, negative refraction remains an unsolved issue and other strategies, in particular based on new mechanism, are extremely desired.

    The parity-time (PT) symmetric systems [17-21] provide a new way. Recently, negative refraction was theoretically demonstrated in a pair of PT symmetric metasurfaces [22]. One is a gain layer, the other is a loss layer, separated by an air gap (see in Fig. 1(b)); both layers meet PT symmetry. When the PT symmetric system is at the exceptional point [23, 24], negative refraction is seen for wave incident with a specific angle from the loss side. The underlying physics lies on the findings that coherent perfect absorber (CPA) and laser modes can occur in the loss and gain media, respectively, with a unidirectional energy flowing from the gain medium to the loss medium (see the black arrow in Fig. 1(b)). In contrast to other solutions, loss-free and wide angle negative refraction can be realized in PT symmetric systems, without bulk metamaterials or nonlinear effects. However, this working principle of negative refraction is inapplicable for wave incident from the gain side, because CPA and laser modes are not supported in the gain and loss media, respectively. Therefore, there is a unidirectional limit in PT symmetric systems.

Recently, as a kind of non-Hermitian materials, conjugate metamaterials (CMs) have drawn much attention [25-28]. The feature of CMs is that both permittivity and permeability are complex conjugates of each other, for instance $\varepsilon = |\varepsilon|\exp(i\alpha)$ and $\mu = |\mu|\exp(-i\alpha)$. By employing a CM slab, perfect lens as a limited case was demonstrated for CMs with $\alpha = \pi$ [25]. In particular, when $\alpha = \pi/2$, such CMs are purely imaginary; CPA and laser modes (even their coexistence) can be found in a slab of purely imaginary conjugate metamaterials (PICMs) [26]. Therefore, the PICMs might be used to achieve negative refraction, overcoming the disadvantages of both bulk metamaterials and PT symmetric systems. The aim of this work is to reveal a new strategy based on PICMs for realizing negative refraction. We will systematically and thoroughly investigate the system of two PICM layers separated by an air gap (see Fig. 1(b)). We show that: (i) for the case of PICMs with CPA and laser modes functioning in different conditions, nonreciprocal perfect negative refraction can be seen (that is, perfect negative refraction is captured for the incident wave from one side; yet for the incident wave from the other side, imperfect negative refraction with scattering will appear). To some extent, such results are beyond the unidirectional limit of negative refraction in PT symmetric systems; (ii) when CPA and laser modes work simultaneously in PICMs, perfect bidirectional negative refraction can be realized only with a pair of identical PICMs, which is a leap forward to the outcomes from PT symmetric systems. (iii) Specially, if the refractive indices of PICMs are less than unity, negative refraction will disappear. Instead, asymmetric electromagnetic (EM) excitation takes place in the air gap, although bidirectional total transmission happens. (iv) What's more, bidirectional negative refraction and planar focusing [29] are well demonstrated in more generalized PICMs with sub-wavelength thicknesses. In fact, negative refraction in PICM systems is also loss-free and wide angle, because it shares the same working principle with that in PT symmetric systems. Therefore, our proposed PICM systems show more fantastic phenomena as well as underlying physics.

## 2. Analytical results

Before further discussions, let us revisit the CPA and laser effects in a PICM slab in air [26]. As shown in Fig. 1(a), the so-called CPA is that two coherent incident waves are totally absorbed in a PICM slab without any reflection (see the blue arrows). In contrast, for the incident waves without coherence, laser mode with two intense outgoing waves will be generated, which is described by the red arrows in Fig. 1(a). Such CPA and laser modes in PICMs also can lead to negative refraction. The schematic diagram of the fundamental principle for realizing negative refraction is illustrated in Fig. 1(b), where two PICM slabs with thicknesses of $d$ are separated by an air gap with a length of $l$. If the left and the right PICM slabs support CPA mode and laser mode, respectively, the backward energy flow (see the black arrow in Fig. 1(b)) can take place in the air gap for wave incident from the left side, bringing about the effect of negative refraction. As the loss and gain portions are simultaneously included in PICMs, accordingly, the right and left PICM slabs can also support CPA and laser modes, respectively. Likewise, negative refraction can happen for the right incidence, and bidirectional negative refraction could be realized in our PICM system.

To realize negative refraction in a PICM system, the condition of CPA and laser modes must be addressed, which can be obtained by analyzing the wave scattering of a PICM slab. We first study a single PICM slab (PICM-1) whose parameters are $\varepsilon_1 = -i\varepsilon$ and $\mu_1 = i\mu$, where $\varepsilon$ and $\mu$ are positive numbers. We take the TE (transverse-electric) polarized wave with the electric field only along $z$ direction under consideration. After eigen modes analysis, the dispersion relationships for both CPA and laser modes are [26],

$$\eta_1 = -i\sigma \cot(k_{1x}d/2) \text{ (odd)}, \tag{1a}$$

$$\eta_1 = i\sigma \tan(k_{1x}d/2) \text{ (even)}, \tag{1b}$$

where $\eta_1 = k_x\mu_1/k_{1x}$, with $k_x = (k_0^2 - \beta^2)^{1/2}$ and $k_{1x} = (n_1^2 k_0^2 - \beta^2)^{1/2}$ ($n_1^2 = n^2 = \varepsilon\mu$); $k_0$ and $\beta$ are the wave vector in air and the *y*-component of propagating wave vector (tangential momentum), respectively; $\sigma = 1$ for laser, while $\sigma = -1$ for CPA. Even modes are defined as symmetric modes, *i.e.*, the field distribution in *x* direction is symmetric, and odd modes are anti-symmetric modes, *i.e.*, the field distribution in *x* direction is anti-symmetric. Note that CPA mode is the time-reversed counterpart of laser mode, which is applicable in PICMs. Therefore, if there is another PICM slab with $\varepsilon_2 = i\varepsilon$ and $\mu_2 = -i\mu$ (PICM-2), we can also find the CPA and laser modes. As PICM-2 is the time-reversed form of PICM-1, for a single PICM-2 slab, $\sigma = 1$ and $\sigma = -1$ in Eq. (1) are related to CPA and laser modes, respectively. In other words, if we obtain CPA (laser) modes in a PICM-1 slab, accordingly, laser (CPA) modes will be achieved in a PICM-2 slab.

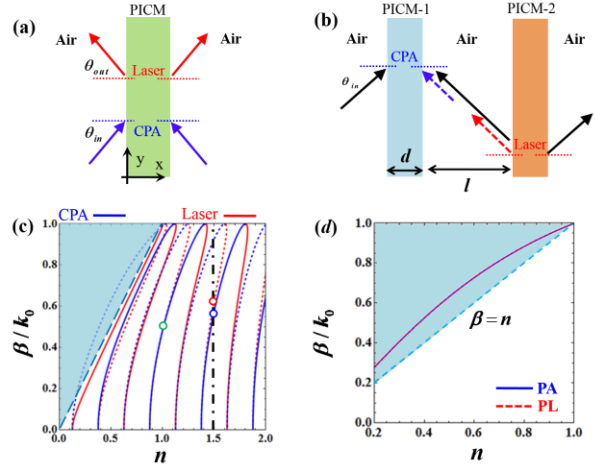

Fig. 1. (a) A schematic diagram of CPA and laser modes in a PICM slab. (b) A schematic diagram of negative refraction using PICMs. (c) is the dispersion relationship ($\beta$ vs $n$) of CPA modes (the blue curves) and laser modes (the red curves) for PICM-1. The corresponding solid and dashed curves are even and odd modes, respectively. The dashed line inside the color region is the critical angle for a PICM slab. The green point denotes the coexistence of CPA and laser modes for $n=1$; the blue and red points represent that CPA and laser modes happen at different tangential momenta for $n=1.5$; (d) is the dispersion relationship ($\beta$ vs $n$) of PA modes (the blue solid curve) for PICM-1 and PL modes (the red dashed curves) for PICM-2. In all the calculations $d=2\lambda$.

Based on Eq. (1), Fig. 1 (c) analytically shows the dispersion relationships $\beta(n)$ for both CPA modes (the blue curves) and laser modes (the red curves) for the PICM-1 slab with $d=2\lambda$ ($\lambda$ is the working wavelength), in which the solid and dashed curves are corresponding to the even and odd modes, respectively. While for PICMs-2, the red and blue curves in Fig. 1(c) are corresponding to CPA and laser modes, respectively. Note that the studies about the CPA and laser modes in PICM-1 and their associated physics have been revealed in Ref. [26]. Therefore, we directly use the relevant results and focus on how to realize the negative refraction effect based on PICMs. Several main points are required to recall for further discussions. (1) for a PICM slab of interest, it can support not only CPA mode, but also laser mode. Particularly, both CPA and laser modes can happen at the same $\beta(n)$ (e.g., $n=1$, see the green point in Fig. 1(c)) or at different $\beta(n)$ (e.g., $n=1.5$, see the red and blue points in Fig. 1(c)), which is determined by the resonance conditions of CPA and laser modes [25]. (2) Specifically, for a lower refractive index ($0<n<1$), conventionally there is a critical angle $\theta_c = \arcsin(n)$ for total internal reflection. The critical angle for a PICM slab marked by the dashed line inside the color region is shown in Fig. 1(c). Specially, only the CPA mode or laser mode for PICMs can occur with the wave vector $\beta(n)$ beyond its critical angle. Concretely, the PICM-1 slab only can support CPA mode, and such two coherent incoming waves for CPA mode in the PICM-1 slab can be simplified as the case of a single incoming wave perfectly absorbed by it. As the coherent conditions are not required, we defined such modes as perfect absorber (PA) modes with dispersion relationship given as,

$$\eta_1 = k_x \mu_1 / k_{1x} = 1, \text{ with } k_{1x} = i\sqrt{\beta^2 - n_1^2 k_0^2} . \tag{2}$$

While for a PICM-2 slab with a lower refractive index, only laser mode can happen in the case of the required momentum beyond the critical angle. In this case, opposite to PA mode, a single incoming wave will be greatly enhanced in a PICM-2 slab without any reflection, which is defined as a perfect laser (PL) mode, and the corresponding dispersion is,

$$\eta_2 = k_x \mu_2 / k_{2x} = -1, \text{ with } k_{2x} = i\sqrt{\beta^2 - n_2^2 k_0^2} . \tag{3}$$

Figure 1(d) shows these PA modes supported by PICM-1 and PL modes supported by PICM-2; they exactly coincide together. By observing the dispersion relationships in Figs.1(c) and 1(d), Table 1 summarizes all the possible CPA and laser modes in PICMs, classified into three cases. The first one is that CPA and laser modes happen at different conditions (see the green area in Table 1), i.e., with different $\beta$. The second one is that CPA and laser modes happen simultaneously (see the orange area in Table 1), i.e., with the same $\beta$. The third one is that PA mode or PL mode occurs in a PICM slab with a lower refractive index for the incident angle beyond its critical angle (see the blue area in Table 1). Such cases will be successively employed to study the scenario in Fig. 1(b).

| Index($n$) | $0<n<1$ | | | | $n=1$ | | $n>1$ | |
|---|---|---|---|---|---|---|---|---|
| Modes / Materials | $\beta=0$ | $\beta \neq 0$ | | | $\beta \neq 0$ | $\beta=0$ | $\beta \neq 0$ | |
| | | $\beta_1$ | $\beta_2$ | $\theta > \theta_c$ | | | $\beta_1$ | $\beta_2$ |
| PICM-1 | CPA & Laser | CPA | Laser | PA | CPA & Laser | CPA & Laser | CPA | Laser |
| PICM-2 | CPA & Laser | Laser | CPA | PL | CPA & Laser | CPA & Laser | Laser | CPA |

Table 1. The CPA and laser modes in PICMs with different parameters for TE polarization

## 3. Numerical demonstration for negative refraction in PICMs

Firstly, we illustrate the negative refraction effect for the case that the CPA and laser modes happen at different conditions. We take $n$=1.5 for example. Both the PICM-1 and PICM-2 slabs have the same thicknesses of $d=2\lambda$, which are placed in the left and right sides of the air gap with $l=3\lambda$. Figure. 1(c) tells us that for PICM-1 slab, $\beta_{cpa}^{(1)}$=0.577$k_0$ is chosen for CPA mode, and $\beta_{laser}^{(1)}$=0.627$k_0$ is chosen for laser mode; for PICM-2 slab, $\beta_{laser}^{(2)}$=0.577$k_0$ is for obtaining laser mode, and $\beta_{cpa}^{(2)}$=0.627$k_0$ is for realizing CPA mode. Now we consider a TE plane wave strikes this system from air. For the left incidence with $\theta = 35.20°$ (*i.e.*, $\beta_{in}$=0.577$k_0$), the PICM-1 slab and PICM-2 slab can exactly function as CPA and laser modes, respectively. As a result, perfect negative refraction with backward energy flow (see the black arrows) can be realized in the air gap, which is numerically verified in Fig. 2(a). While for wave incident with $\theta = 35.20°$ from the right side, as it does not exactly match the required incident angle $\theta=38.80°$ ($\beta_{in}=0.627k_0$) of obtaining CPA for PICM-2, yet close to such an incident angle, PICM-2 and PICM-1 work as imperfect CPA and laser modes, respectively. Consequently, negative refraction with some scattering happens as shown in Fig. 2(b), where the backward energy flow still occurs in the air gap. Likewise, for the incident waves with $\theta=38.80°$, negative refraction with some scattering will appear from wave incident from the left side (see Fig. 2(c)) and perfect negative refraction can be observed for the right incidence (see Fig. 2(d)). Therefore, for PICMs with CPA and laser modes occurring in different conditions, perfect negative refraction can take place for wave coming from one side, while negative refraction with some scattering appears for wave incident from the other side. Such scattering is proportional to the momentum deviation of CPA and laser modes in PICMs.

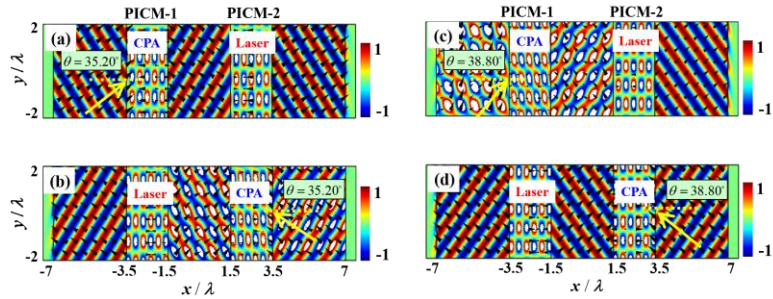

Fig. 2. The simulated electric field patterns for negative refraction using a pair of PICMs with CPA and laser modes occurring in different conditions. (a) and (b) are the simulated electric field patterns for the incident waves with $\theta=35.20°$ from the left and right sides, respectively. (c) and (d) are the simulated electric field patterns for the incident waves with $\theta=38.80°$ from the left and right sides, respectively. In all plots, $\varepsilon=\mu=n=1.5$ is set for both PICM-1 and PICM-2. The thickness of both PICM-1 and PICM-2 are $d=2\lambda$, and the distance between these two PICMs is $l=3\lambda$.

Secondly, we discuss the negative refraction for PICMs with CPA and laser modes happening simultaneously. In such a condition, only two identical PICMs are required to realize negative refraction, as CPA and laser modes coexist in PICMs. Alternatively, we choose PICM-1 to explore negative refraction, and similar effect can be obtained by using PICM-2. Based on Eq. (1), we can derive the coexistent conditions for CPA and laser modes in PICM-1 slab [26], which is $\eta_1 = k_x\mu_1/k_{1x}=i$. After some simplifications, such a condition can be further given as: $\beta_{cpa/laser}=0$ or $n_{cpa/laser}=1$ (see the orange area in Table 1). For the case of $\beta_{cpa/laser}=0$, there exist a series of $n_{cpa/laser}$ to obtain the simultaneous realization of CPA and laser modes, which is given by $n_{cpa/laser}k_0 d=(\pi/2+N\pi)$ [25], where $N$ is an positive integer; for the case of $n_{cpa/laser}=1$, there are a series of $\beta_{cpa/laser}\neq 0$ for the coexistence of CPA and laser modes, which can be deduced from $(1-\beta_{cpa/laser}^2)^{1/2}k_0 d=(\pi/2+N\pi)$ [25, 26]. To display negative refraction clearly, we choose the case of $n_{cpa/laser}=1$ and consider two identical PICM-1 slabs in Fig. 1(b). As shown by Fig. 1(c), for the PICM-1 slab with $d=2\lambda$, one solution for obtaining CPA and laser modes concurrently is $\beta_{cpa/laser}=0.485k_0$. To match such a wavevector, a TE plane wave with $\theta=29.0°$ is incident from air into PICM-1 system, as shown in Figs. 3(a) and 3(b), where perfect bidirectional negative refraction happens with the backward energy flow appearing in the air gap. Therefore, when CPA and laser modes function simultaneously in a PICM slab, perfect bidirectional negative refraction can be realized by using a pair of identical PICMs. Different from these in Fig. 2 which meet PT symmetry, here the left and right PICMs are identical and symmetric, wave propagation is reciprocal.

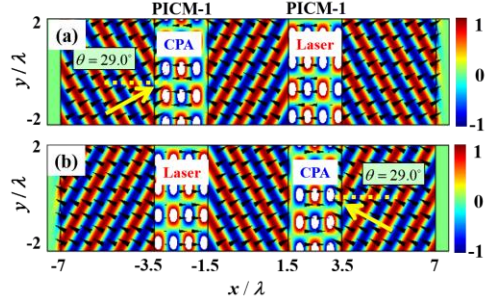

Fig. 3. The simulated electric field patterns for negative refraction using PICMs-1 with CPA and laser modes occurring at the same condition. (a) and (b) are the simulated electric field patterns for incident waves with $\theta=29°$ from the left and right sides, respectively. $n=1$ is set for PICM-1.

Thirdly, we will investigate negative refraction for PA mode and PL mode in PICMs with a lower refractive index. Without loss of generality, $n=0.5$ is chosen for PICMs to achieve PA and PL modes. Based on Eq. (2) and Eq. (3), PA mode for PICM-1 and PL mode for PICM-2 are obtained by sharing the same tangential momenta, i.e., $\beta_{pa/pl}=0.6847k_0$. For wave incident with $\theta=39.23°$ (it exactly matches $\beta_{pa/pl}=0.6847k_0$) from the left side, it experiences a dissipative process in PICM-1 without any reflection (PA mode works in PICM-1). Afterwards, the transmitted wave with near zero amplitude in the air gap is enhanced symmetrically in PICM-2 (PL mode works in PICM-2). Such a dissipative and enhanced process is well depicted in Fig. 4(a). Furthermore, as shown by the insert in Fig. 4(a), the negative refraction phenomenon is disappeared. For the right incidence with $\theta=39.23°$, it will perfectly undergo an enhanced and then dissipative process in PICM-2 and PICM-1 respectively (see Fig. 4(b)). Negative refraction with the backward energy flow also disappears, with extremely enhanced field in the air gap (see the insert in Fig. 4(b)). Therefore, for PICMs with a lower refractive index, negative refraction will disappear for both incident directions. Instead, there is asymmetric field amplitude in the air gap. Physically, it stems from the disappearance of CPA and laser modes, so that the channel of backward energy flow cannot be established. In addition, as the dissipative and amplified coefficients of PA and PL modes are the same, perfect total transmissions without any reflections for both incident directions happen.

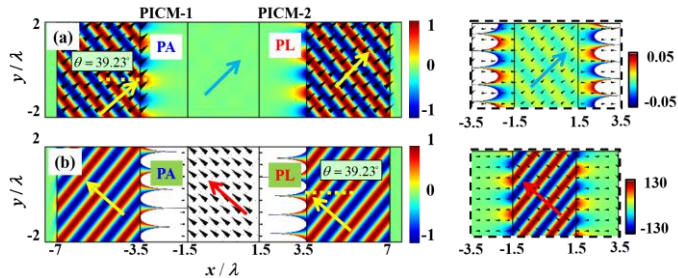

Fig. 4. The simulated electric field patterns for PA and PL modes in PICMs with a lower refractive index. (a) and (b) are the simulated electric field patterns for the incident waves with $\theta=39.23°$ from the left and right sides, respectively. $n=0.5$ is set for both PICMs. The inserts are the related field patterns from $x=-3.5\lambda$ to $x=3.5\lambda$ ($y=-2\lambda$ to $y=2\lambda$) and the arrows in the inserts denote the directions of energy flow in the air gaps.

## 4. Negative refraction and planar focusing in generalized PICMs with a sub-wavelength thickness

In the above discussions, as a demonstration of fundamental principle, PICMs with a larger thickness ($d=2\lambda$) are employed to realize some interesting phenomena. In fact, even for PICMs with a sub-wavelength thickness, except the case of PICMs with lower refractive index, the above similar results could be obtained as well, yet at the price of a higher $n$ in PICMs. As we have demonstrated previously that CPA and laser modes with $\beta_{cpa/laser} \neq 0$ only simultaneously happen for PICMs with $n=1$, therefore the concurrent condition of CPA and laser modes with $\beta_{cpa/laser} \neq 0$ is not available for PICMs with a sub-wavelength thickness. Consequently, negative refraction could not be realized with a pair of identical sub-wavelength PICM slabs. To solve this problem, we extend the concept of PICMs to a more generalized case, i.e., purely imaginary metamaterials (PIMs). For instance, PIM-1 is the general case of PICM-1, because there is no need for $\varepsilon=\mu$ in PIM-1 ($\varepsilon_1 = -i\varepsilon$ and $\mu_1 = i\mu$), which can release a parameter freedom for Eq. (1). Thereupon, CPA and laser modes with $\beta_{cpa/laser} \neq 0$ might be realized simultaneously in a sub-wavelength PIM-1 slab. For example, the sub-wavelength scale is assumed as $d=0.1\lambda$ and $\beta_{cpa/laser} = 0.5k_0$ is chosen for realizing CPA and laser modes concurrently. By analyzing Eq. (1), $\varepsilon=2.253$ and $\mu=2.88$ are one of the several solutions to realize the coexistence of CPA and laser modes in PIM-1. Here we use such PIM-1 to verify negative refraction based on the structure in Fig. 1(b), where the distance between PIM-1 slabs is $l=3\lambda$. The corresponding electric field patterns for both left and right incidences are respectively shown in Figs. 5(a) and 5(b), where perfect negative refractions with the backward energy flow happen for both cases. When other appropriate parameters are applied in PIM-1 slabs, perfect bidirectional negative refraction with the desired refractive angle could be realized as well. Therefore, if we engineer well-defined profiles in PIM-1 slabs to obtain CPA and laser modes simultaneously, wide angle negative refraction can be achieved for planar focusing [22, 29]. For example, the focus length is set as $f=6\lambda$, and a pair of PIM-1 slabs with $d=0.1\lambda$ are placed at $x=-6\lambda$ and $x=6\lambda$ respectively. For each slab with a length of $15\lambda$, it is divided into 15 cells, and each with a length of $\lambda$ is respectively marked from $N=-7$ to $N=7$ with position from bottom to top. Accordingly, for realizing CPA and laser modes simultaneously in each cell, the required tangential momentum is given as $\beta_N = a\tan(|N\lambda|/f)$. As a result, the well-defined profiles of PIM-1 in each cell can be obtained by employing Eq. (1) and $\beta_N$. By using these profiles in a pair of PIM-1 slabs, the corresponding simulated electric field patterns for planar focusing are shown in Figs. 5(c) and 5(d), where the point sources are located at $(-12\lambda, 0)$ and $(12\lambda, 0)$ respectively. Apparently, there are two focusing points located at $(0, 0)$ and $(12\lambda, 0)$, caused by the wide angle negative refraction for the point source at $(-12\lambda, 0)$ (see Fig. 5(c)). Likewise, there are two focusing points at $(0, 0)$ and $(-12\lambda, 0)$ for the point source at $(12\lambda, 0)$ (see Fig. 5(d)). As these cells can support CPA and laser modes simultaneously for the point sources in these two particular positions, bidirectional planar focusing can be realized, which is not accessible in PT symmetric systems [22, 29]. In addition, if the point source deviates slightly from these two locations ($\pm 12\lambda, 0$), e.g., the point source is placed at $(11\lambda, \lambda)$, as wide angle negative refraction still works, the similar planar focusing could be observed. Note that the planar focusing from PT symmetric system has the diffraction-limited imaging resolution, because only the propagating wave components can be focused [29]. Therefore, the image resolutions in our PIM-1 system also don't break through the diffraction limit, as the working principle in PIM-1 system is the same with that in PT symmetric system.

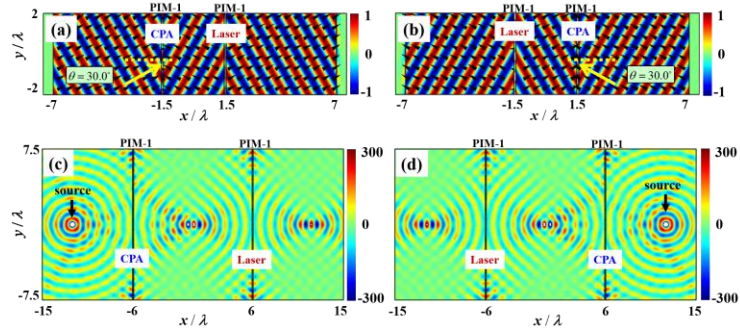

Fig. 5. The simulated electric field patterns for sub-wavelength PIM-1 with CPA and laser modes occurring simultaneously. (a) and (b) are the simulated electric field patterns of negative refraction for incident waves with $\theta=30°$ from the left and right sides, respectively. $\varepsilon=2.253$ and $\mu=2.88$ are set for PIM-1 ($\varepsilon_1=-i\varepsilon$ and $\mu_1=i\mu$). (c) and (d) are the simulated electric field patterns of planar focusing for the point sources locating at ($-12\lambda$, 0) and ($12\lambda$, 0) respectively, where a pair of PIM-1 slabs are equipped with the same well-defined profiles: $\varepsilon=2.5$ and $\mu=2.5$ for cells with $N=0$; $\varepsilon=2.483$ and $\mu=2.534$ for cells with $N=\pm1$; $\varepsilon=2.414$ and $\mu=2.65$ for cells with $N=\pm2$; $\varepsilon=2.287$ and $\mu=2.822$ for cells with $N=\pm3$; $\varepsilon=2.126$ and $\mu=3.1$ for cells with $N=\pm4$; $\varepsilon=1.931$ and $\mu=3.477$ for cells with $N=\pm5$; $\varepsilon=1.7$ and $\mu=4.04$ for cells with $N=\pm6$; $\varepsilon=1.425$ and $\mu=4.937$ for cells with $N=\pm7$.

## 5. Discussion and conclusion

In conclusion, bidirectional negative refraction and planar focusing have been well demonstrated in the proposed PICM systems, even with a sub-wavelength thickness. The underlying physics lies in the coexistence of CPA and laser modes in PICMs. Therefore, negative refraction in our PICM systems overcomes the disadvantages of the current technologies [22, 29], with loss-free and bidirectional features revealed. Specially, in a pair of PICMs with a lower refractive index, asymmetric field amplitude with perfect total transmission can be observed. In fact, for transverse magnetic (TM) polarization (the magnetic field is along $z$ direction), all the similar results could be obtained. Consequently, negative refraction and asymmetric field amplitude can be realized without limitation of polarizations. Therefore, our proposed PICM system reveals more fantastic phenomena, and the generalized PICMs are also worthy of exploring further.

## Funding

This work was supported by the National Natural Science Foundation of China (grant No. 11604229), the National Science Foundation of China for Excellent Young Scientists (grant no. 61322504), the Postdoctoral Science Foundation of China (grant no. 2015M580456), and the Fundamental Research Funds for the Central Universities (Grant No. 20720170015).